\documentclass[pra,twocolumn,showpacs,preprintnumbers,amsmath,amssymb,tightenlines,epsfig]{revtex4}

\usepackage{graphicx}% Include figure files
\usepackage{dcolumn}% Align table columns on decimal point
\usepackage{bm}% bold math
\usepackage{epsfig}

\def\hamax{ \hat{H}_{0a}(\mathbf{x})}
\def\hamaxs{\hat{H}_{0a}(\mathbf{x'})}
\def\psiax{\hat{\Psi}_{a}(\mathbf{x})}

\def\psimx{\hat{\Psi}_{m}(\mathbf{x})}

\def\psiadax{\hat{\Psi}^{\dagger}_{a}(\mathbf{x})}

\def\psiadmx{\hat{\Psi}^{\dagger}_{m}(\mathbf{x})}

\def\phiar{\tilde{\phi}_{a}(R)}
\def\phiard{\tilde{\phi}_{a}(R')}
\def\phiarstar{\tilde{\phi}_{a}^{*}(R)}

\def\phiax{\phi_{a}(\mathbf{x})}
\def\phiaxs{\phi_{a}(\mathbf{x'})}
\def\phimx{\phi_{m}(\mathbf{x})}

\def\phicax{\phi^{*}_{a}(\mathbf{x})}
\def\phicaxs{\phi^{*}_{a}(\mathbf{x'})}

\def\gnx{G_{N}\left( \mathbf{x},\mathbf{x'} \right)}
\def\gnxstar{G_{N}^{*}\left( \mathbf{x},\mathbf{x'} \right)}

\def\gax{G_{A}\left( \mathbf{x},\mathbf{x'} \right)}
\def\gaxstar{G_{A}^{*}\left( \mathbf{x},\mathbf{x'} \right)}

\def\gnxx{G_{N}\left( \mathbf{x},\mathbf{x} \right)}
\def\gaxx{G_{A}\left( \mathbf{x},\mathbf{x} \right)}
\def\gapol{\tilde{G}_{A}\left(R,R', \beta \right)}
\def\gnpol{\tilde{G}_{N}\left(R,R',\beta \right)}
\newcommand{\gapolm}[1]{\tilde{G}_{A}^{(#1)}\left(R,R' \right)}
\newcommand{\gnpolm}[1]{\tilde{G}_{N}^{(#1)}\left(R,R' \right)}
\newcommand{\gapolmstar}[1]{\tilde{G}_{A}^{(#1)*}\left(R,R' \right)}
\newcommand{\gnpolmstar}[1]{\tilde{G}_{N}^{(#1)*}\left(R,R' \right)}
\def\gadiag{\bar{G}_{A}\left(R \right)}
\def\gndiag{\bar{G}_{N}\left(R  \right)}
\def\gadiags{\bar{G}_{A}\left(R' \right)}
\def\gadiagsstar{\bar{G}_{A}^{*}\left(R' \right)}
\def\gndiags{\bar{G}_{N}\left(R'  \right)}

 \newcommand{\intas}{\int d \mathbf{x}\:}
\newcommand{\mods}[1]{\left| #1 \right| ^{2}}
\newcommand{\deltf}{\delta^{(3)}(\mathbf{x}-\mathbf{x}')}
\newcommand{\figuresize}{\columnwidth}
\newcommand{\sub}[2]{{#1}_{\mbox{\!\! \scriptsize #2}}}

\def\beq{\begin{equation}}
\def\eeq{\end{equation}}
\def\nnl{\\[0.15cm] \nonumber}
\def\nl{\\[0.15cm] }
\def\CR{\nonumber\\[0.15cm]}
\newcommand{\rref}[1]{Ref.~\cite{#1}}
\newcommand{\fref}[1]{Fig.~\ref{#1}}
\newcommand{\eref}[1]{Eq.~(\ref{#1})}
\newcommand{\sref}[1]{section \ref{#1}}
\begin{document}

%\preprint{Version: In progress, \today}
\title{Collapsing Bose-Einstein condensates beyond the Gross-Pitaevskii approximation}
\author{S. W\"uster, J.J. Hope and C.M. Savage}
\affiliation{ARC Centre of Excellence for Quantum-Atom Optics, Department of Physics, 
Australian National University, Canberra
ACT 0200, Australia }
\email{sebastian.wuester@anu.edu.au}

\begin{abstract}
We analyse quantum field models of the bosenova experiment, in which $^{85}$Rb Bose-Einstein condensates were made to collapse by switching their atomic interactions from repulsive to attractive. Specifically, we couple the lowest order quantum field correlation functions to the Gross-Pitaevskii function, and solve the resulting dynamical system numerically. Comparing the computed collapse times with the experimental measurements, we find that the calculated times are much larger than the measured values. The addition of quantum field corrections does not noticeably improve the agreement compared to a pure Gross-Pitaevskii theory.
\end{abstract}

\pacs{03.75.Kk, 03.75.Nt, 03.70.+k}
\maketitle

\section{Introduction}

Dilute gas Bose-Einstein condensates (BECs) display not only interesting classical nonlinear field physics, but also quantum field physics, such as the Mott insulator transition \cite{greiner:mott}. These BECs are especially interesting because the weak atomic interactions make the physics sufficiently simple that we expect to get quantitative agreement between experiments and theory. In this paper we analyse one of the more straightforward aspects of the JILA bosenova experiment \cite{jila:nova}: the time to collapse. It has previously been shown that the Gross-Pitaevskii (GP) theory substantially overestimates these collapse times \cite{savage:coll}. We show that adding the lowest order quantum field corrections does not improve the situation. Thus an open question remains: what explains the discrepancy between theory and experiment?

The collapse experiments at JILA \cite{jila:roberts,jila:nova} represent the first completely controlled studies of unstable BECs. Earlier experiments on unstable $^{7}$Li condensates \cite{gerton:lithium,bradley:lithium} could not provide such a well defined initial state for the collapse since the condensate was continuously fed by a thermal cloud and collapse was therefore triggered in a random fashion. Nonetheless these were the first experiments to verify theoretical predictions about the critical atom number in attractive BECs. In the experiment described in \rref{jila:nova}, a stable  $^{85}$Rb condensate was prepared with vanishing interactions, and then the scattering length was switched to a negative (attractive) value using a Feshbach resonance. 

The experiment demonstrated a number of complicated and interesting features of the condensate collapse, which was controlled via the use of a Feshbach resonance. %Numerous theoretical articles have been published in order to explain the outcome of the experiment 
We have extended existing theoretical studies \cite{adhikari:coll,adhikari:jets,saito_ueda:coll,saito_ueda:intermitt,saito_ueda:powerlaw,saito_ueda:picture,shlyapnikov:coll,savage:coll,bao:jets,duinestoof:feshbachdynamics, metens,calz:hu,duinestoof:evaporation,holland:burst} by applying %However, we consider the discrepancy between measured collapse times and existing, theoretical results based on the GP equation, as a case where a qualitative explanation has been found, but quantitative agreement has not yet been achieved. We will report our investigation of a possible source for this discrepancy: quantum field corrections to the mean field dynamics. 
%Extending these studies, we employed an extension of Gross-Pitaevskii mean field theory, 
the dynamical Hartree-Fock Bogoliubov (HFB) equations with realistic experimental parameters. 

%The results of this experiment presented multiple challenges to theory. It takes a delay time $\sub{t}{collapse}$ until the peak density increases sufficiently for inelastic processes to significantly reduce the number of atoms in the condensate.
%, which depends on the strength of the attractive interaction $\sub{a}{collapse}$. 
%Bursts of hot atoms were also produced. 
%in each collapse and the ejection of  comparatively cold atom jets in the plane normal to the condensate symmetry axis. 

This paper is organized as follows: In the first part we will give an overview of the JILA experiment and the associated theoretical literature.
%, focusing on the dependence of the collapse time $\sub{t}{collapse}$ on the scattering length $\sub{a}{collapse}$. 
We then introduce our quantum field models, and finally present the results of our simulations. 

\subsection{Experiment}

%At first we will give a more detailed description of the experimental results from \rref{jila:nova} than presented in the introduction. It was found that 
The collapsing condensate was observed to lose atoms until the atom number reduced to about the critical value below which a stable condensate can exist \cite{jila:nova}. The dependence of the number of remaining atoms on time since initiation of the collapse $\sub{\tau}{evolve}$ was measured for the case of an initial state with $\sub{N}{init}=16000$ atoms and  repulsive interaction corresponding to $\sub{a}{init}=+7a_{0}$, where $a_{0}$ is the hydrogen Bohr radius. 
The onset of number loss is quite sudden, with milliseconds of very little loss followed by a rapid decay of condensate population (within ~0.5 ms) after which the condensate stabilizes again. This behavior results from the scaling of the loss rate with the cube of the density, the peak value of which rises as $1/(\sub{t}{collapse}-t)$ near the collapse point \cite{shlyapnikov:coll}. This allows a precise definition of the collapse time $\sub{t}{collapse}$, the time after initiation of the collapse up to which only negligible numbers of atoms are lost from the condensate. Another quantitative result of the experiment is the dependence of $\sub{t}{collapse}$ on the magnitude of the attractive interaction that causes the collapse, parametrised by the (negative) scattering length $\sub{a}{collapse}$. These measurements are performed from an initial state with $\sub{N}{init}=6000$ atoms in an ideal gas state (with interaction between them tuned to zero). 
The $\sub{t}{collapse}$ datapoints presented in the original paper have undergone one revision of their $\sub{a}{collapse}$ values by a factor of 1.166(8) due to a more precisely determined background scattering length \cite{jila:revision}. 
Although the main focus of this paper shall be on the collapse time, we mention two other striking features of the experiment: the appearance of 'bursts' and 'jets'. One fraction of the atoms that are lost during the collapse is expelled from the condensate at quite high energies ($\sim$100 nK to $\sim$400~nK, while the condensate temperature is 3 nK); this phenomenon was referred to as 'bursts'. 
%The energies and the anisotropy between radial and axial direction of these burst atoms have been determined for a variety of values $\sub{a}{collapse}$ and $\sub{N}{init}$. 
Finally, when the collapse was interrupted during the period of number loss by a sudden jump in the scattering length, another atom ejection mechanism was observed: 'jets' of atoms emerge, almost purely in the radial direction and with temperatures a lot lower than that of the bursts (a few nK). 
In addition to the two initial scenarios described above there have been unpublished measurements on a third initial state with $\sub{N}{init}=4000$ and $\sub{a}{init}=+2500 a_{0}$ \cite{unpubl}. To simplify reference to the three different initial states for the collapse, we label them as in Tab. \ref{inistateslist}.
% \newcounter{Ccount}
%  \begin{quote}
% \begin{list}{case (\roman{Ccount})}
%    {\usecounter{Ccount}}
%    \setlength{\labelwidth}{3cm}
%  \item  \hspace{1cm}  $\sub{N}{init}=16000$ and $\sub{a}{init}=+7a_{0}$,
%  \item  \hspace{1cm}  $\sub{N}{init}=6000$   and $\sub{a}{init}=0$,
%  \item  \hspace{1cm}  $\sub{N}{init}=4000$   and $\sub{a}{init}=+2500a_{0}$.
%  \end{list}     
%  \end{quote}
\begin{table}
\vskip 0.5cm
\begin{center}
\begin{tabular}{|ccc|}
\cline{1-3}
         & $N_{init}$ & $a_{init}$\\
\hline
case (i) & $16000$ & $+7a_{0}$\\
case (ii) & $6000$ & $0$\\
case (iii) & $4000$ & $+2500a_{0}$\\
\cline{1-3}
\end{tabular}

\end{center}
\caption{Initial states for different collapse scenarios.}
\label{inistateslist}
\end{table}

%%
%\vskip 0.5cm
%\begin{center}
%%
%\begin{tabular}{||l||c|c||}
%\cline{1-3}
%         & $N_{init}$ & $a_{init}$\\
%\hline\hline
%case (i) & $16000$ & $+7a_{0}$\\
%case (ii) & $6000$ & $0$\\
%case (iii) & $4000$ & $+2500a_{0}$\\
%\cline{1-3}
%\end{tabular}
%\end{center}
%

\subsection{Theory}
\subsubsection{Collapse time}
\begin{figure}
%:
\epsfig{file={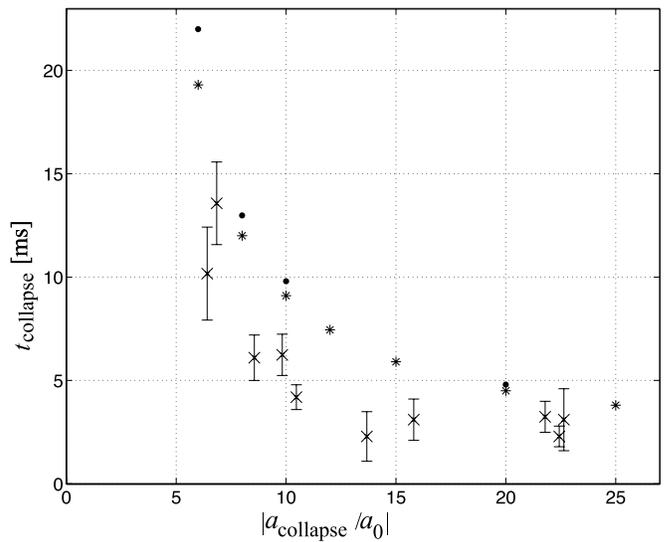},width=\figuresize}
\caption{Experimental and numerical results for the collapse time $t_{\rm{collapse}}$ versus scattering length $a_{\rm{collapse}}$ after a switch from $a = 0$ to $a_{\rm{collapse}}$ in case (ii). The experimental points ($\times$) and their error bars are from Fig.~2 of \cite{jila:nova} taking into account the revision of scattering length values in \cite{jila:revision}. Numerical results are given for exact, cylindrical geometry ($\bullet$) and for spherical geometry ($\ast$) with trap frequency equal to the geometric mean of the experimental values. $K_3 = 1 \times 10^{-27}$ cm$^6$s$^{-1}$.}
\label{fig:collapsetime}
\end{figure}

The unstable attractive BEC and its loss of atoms during collapse can be modeled numerically with the Gross-Pitaevskii equation (GPE) if a phenomenological three body loss term is included. Numerical solutions to this equation for exact experimental parameters and geometry have been reported by Adhikari \cite{adhikari:coll,adhikari:jets}, Saito and Ueda \cite{saito_ueda:coll,saito_ueda:intermitt,saito_ueda:powerlaw,saito_ueda:picture}, Santos and Shlyapnikov \cite{shlyapnikov:coll}, Savage, Robins and Hope \cite{savage:coll} as well as Bao, Jaksch and Markovich \cite{bao:jets}. 
Where they consider the initial situation of case (ii), these solutions agree on collapse times which qualitatively describe the experiment but systematically exceed the experimental values. Note that while the remnant atom number after collapse depends strongly on the three-body recombination rate $K_{3}$, which is not well determined near the Feshbach resonance, and for which different values were employed in the references above, the collapse time in case (ii) does not vary much for experimentally reasonable values of $K_{3}$ \cite{savage:coll}. \fref{fig:collapsetime} shows a comparison between GP theory ($\bullet$) and the revised experimental data ($\times$) for case (ii). Also shown are the GP results for spherical symmetry, for comparison with our quantum field theory calculations. The change in $\sub{a}{collapse}$ by a factor of 1.166(8) due to the revision has been included and moved the experimental and theoretical points closer together but they still do not agree. 
%In particular for values $\sub{a}{collapse}/a_{o}\sim 12$ the disagreement in $\sub{t}{collapse}$ is by more than a factor of two. 

In \cite{saito_ueda:coll,shlyapnikov:coll,bao:jets} this situation is accepted as agreement between theory and experiment. But if the GP equation indeed contains all the essential physics to describe $\sub{t}{collapse}$, one should expect better agreement since all crucial parameters of the model are experimentally fixed. In addition there exists another case where the GPE can apparently not account for $\sub{t}{collapse}$ accurately, and that is the above mentioned case (iii) \cite{duinestoof:feshbachdynamics}. One reason for the controversy in the literature might be the fact that the GPE describes collapse times and remnant numbers well for case (i) \cite{savage:coll,adhikari:coll,bao:jets,saito_ueda:coll}. The condensates for cases (i) and (ii) are similar in peak density, ruling out a simple qualitative difference of the densities involved as an explanation for this distinct behaviour.

Complementary to the numerical approach to the problem, one can attempt to find an analytic approximation that describes the essential features of the BEC collapse.  Metens \textit{et al.} \cite{metens} employ a variational approach assuming a Gaussian condensate shape. For values $||a|-|a_{cr}||/|a_{cr}|<<1$, where $a_{cr}$ is the critical scattering length above which the BEC does not collapse, they obtain $\sub{t}{collapse} \sim (|a/a_{cr}| -1)^{-1/4}$, in agreement with \cite{saito_ueda:powerlaw}. But this parameter regime does not apply to the experiment, and it turns out that the scaling law becomes modified to $\sub{t}{collapse} \sim (|a/a_{cr}| -1)^{-1/2}$ for scattering lengths away from the critical value  \cite{metens}. Calzetta \textit{et al.} \cite{calz:hu} derive a scaling law for the collapse time from the growth of unstable perturbations of the atom field. Their result agrees with the one cited above, which seems to capture the shape of the function correctly. However both methods are unable to account for the proportionality constant, as both groups fit this parameter or the initial state to obtain agreement with experimental data. 

In summary, mean field theory can describe the shape of the dependence of $\sub{t}{collapse}$ on $\sub{a}{collapse}$ correctly, but for case (ii) the real collapse seems significantly accelerated compared to these predictions. Naturally, one would suspect deviations from GP theory to be the cause of this phenomenon \cite{savage:coll,calz:hu}. 

Yurovsky \cite{yurovsky:instabilities} investigates quantum effects in homogeneous BECs with attractive interaction analytically, using an approximate solution to the operator equation for the condensate fluctuations. Results from \cite{yurovsky:instabilities} regarding the growth of an initially seeded unstable momentum-mode have been experimentally verified \cite{ketterle:instabilities}. Yurovsky also derives an expression for the time evolution of condensate depletion.
%Investigating this idea further is the purpose of this paper. 

\subsubsection{Jets}
Another prominent feature of the JILA experiment was the appearance of radial atomic jets when the collapse was suddenly interrupted. Numerical simulations were able to reproduce them using the GPE \cite{bao:jets,adhikari:jets,saito_ueda:coll}. The phenomenon can be explained as interference of atom emission from highly localized density spikes that are formed during collapse \cite{saito_ueda:coll}. The simulations seem to reproduce jet-shapes, anisotropy and dependence on $\sub{t}{evolve}$ (Fig. 5 of \cite{jila:nova}).
By contrast Calzetta \textit{et al.} \cite{calz:hu} present a mechanism for jet production that is essentially a quantum effect. Their model reproduces the features of the experimental data for atom numbers in jets (Fig. 6 of \cite{jila:nova}) for the range of validity of the approximations involved.
Regarding this, the numerical simulations in \cite{bao:jets,adhikari:jets} yield a similar (partial) agreement with experiment, but do not agree with each other. Another feature of the jets in the collapse experiment is the existence of large fluctuations in the atom number in jets. Calzetta \textit{et al.} \cite{calz:hu} attribute this to a large number uncertainty in squeezed quantum states, while Bao \textit{et al.} \cite{bao:jets} reproduce these fluctuations by slight variations in the position of the initial state condensate wavefunction. 
The existence of jets can thus be predicted by GP theory alone, but it is not yet clear what role quantum effects play.

\subsubsection{Bursts}
Finally, one has to explain the bursts. These are atoms that remain in a detectable, trapped state during collapse but are found performing large amplitude oscillations in the trap. 
%with very high energies compared to the condensate. 
Numerical solutions of the GPE generally show atoms being ejected from the condensate at the time of collapse. In order to be candidates for the burst atoms in the JILA experiment, they need to have rather high energies $\sim$ 100 nK (compare Fig. 4 of \cite{jila:nova}). The ejected atoms within the condensate gain their energy due to a loss in (negative) mean field energy by three body recombination, which is sensitive to the three body loss rate $K_{3}$. Since $K_{3}$ is experimentally not very well constrained in the parameter regime in question \cite{jila:k3}, it is possible to reproduce the atom burst by matching $K_{3}$ on the energy spectrum \cite{saito_ueda:picture}. But in \rref{shlyapnikov:coll} it is shown, that even if $K_{3}$ is fitted to the energy spectrum for \emph{each} value of $\sub{a}{collapse}$, one cannot obtain simultaneous agreement for cases (i) and (ii) described in \rref{jila:nova}. Furthermore, the remnant atom number after collapse $\sub{N}{remn}$ also depends on $K_3$ and the values for $K_3$ that are necessary to obtain high energy bursts in agreement with the experiment yield $\sub{N}{remn}$ that disagree with the experiment. Calzetta \textit{et al.} \cite{calz:hu} also suggest a quantum treatment of the bursts but do not put forward any quantitative predictions. Milstein, Menotti and Holland \cite{holland:burst} present a model calculation that shows that atom-molecule coupling can yield sufficiently energetic atoms, however they do not examine the experimental parameter regime. In this paper, we will follow the formalism developed by Milstein \textit{et al.} but place a lower emphasis on the molecular field.

An alternate model that describes physics beyond the GPE has been proposed by Duine and Stoof \cite{duinestoof:evaporation,duinestoof:feshbachdynamics} and applied to the collapse case (iii). They suggest elastic collisions between condensate atoms as the origin of the burst, and an additional channel by which the condensate can eject atoms. One of the collision partners gets ejected into the noncondensate in the process, while the other one is stimulated back into the condensate. The model yields results that describe well the experimental data for the case (iii) collapse. But case (ii) cannot be treated with a Gaussian variational ansatz \cite{duinestoof:evaporation}, and it is not clear how to incorporate these elastic collisions into a treatment that is focussed on the backreaction of uncondensed atoms on the condensate. 
We will comment further on the relation to our model in the last section.
%, but has not been applied to the other two cases yet. 

A numerical study that accurately reproduces the bursts \emph{and} all other results of the experiment therefore does not yet exist.

\section{The Hartree-Fock Bogoliubov model}
\subsection{Quantum corrections}

To go beyond mean field theory we start from the usual Hamiltonian for a many body system that interacts by a contact potential of the form $V(\mathbf{x} - \mathbf{x}')=U_{o}\deltf$.
%\delta(\mathbf{x} - \mathbf{x}')$.  
In the following, $\hat{\Psi}_{a}(\mathbf{x},t)$ will denote the field operator that annihilates an atom at position $\mathbf{x}$.
\begin{align}
\hat{H} =& \intas \psiadax \hamax \psiax
\CR
 &+ \frac{U_{o}}{2} \intas \psiadax  \psiadax  \psiax \psiax,
\label{Hamiltonian}
\end{align}
where
\begin{align}
\hamax&= -\frac{\hbar^2}{2m} \nabla_\mathbf{x}^2 +V(\mathbf{x}),
\CR
V(\mathbf{x})&=\frac{1}{2}m ( \omega_{\bot}^{2} x^{2}+ \omega_{\bot}^{2} y^{2}+ \omega_{\|}^{2} z^{2}).
\label{Potential}
\end{align}
Here $m$ is the atomic mass ($1.41 \times 10^{-25}$ kg for $^{85}$Rb), $U_{o}$ the atom-atom coupling constant and $\omega_{\bot}$, $ \omega_{\|}$ denote the trapping frequencies in the transverse and axial directions. Now we derive the Heisenberg equation for the field operator, and subsequently decompose $\hat{\Psi}_{a}(\mathbf{x},t)$ into a condensate part $\phi_{a} (\mathbf{x},t)$ and quantum fluctuations $\hat{\chi}(\mathbf{x},t)$, such that $\hat{\Psi}_{a}=\phi_{a}+\hat{\chi}$ and $\langle \hat{\Psi} \rangle=\phi_{a}$. We will describe the fluctuations in terms of their lowest order correlation functions: the normal density $G_{N}(\mathbf{x},\mathbf{x}') = \langle \hat{\chi}^{\dagger} (\mathbf{x}')\hat{\chi}(\mathbf{x}) \rangle$ and anomalous density $G_{A}(\mathbf{x},\mathbf{x}') = \langle \hat{\chi}(\mathbf{x}') \hat{\chi}(\mathbf{x}) \rangle$. In deriving the dynamical equation for the condensate we will factor the expectation values in accordance with Wick's theorem, therefore assuming that the system is in a Gaussian quantum state (i.e. a coherent state or even a squeezed state) \cite{book:blaizot} and obtain:
\begin{align}
i \hbar  \frac{\partial \phiax}{\partial t}&=\left(-\frac{\hbar^2}{2m} \nabla_\mathbf{x}^2 +V(\mathbf{x}) 
 +U_{o} |\phiax |^{2} \right)  \phiax
\CR
&+ 2 U_{o} \phiax \gnxx +U_{o}\phicax \gaxx.
\label{atom_eqn}
\end{align}
This modified Gross-Pitaevskii equation now contains the interaction of the uncondensed component with the mean field. We also model three-body loss from the condensate phenomenologically, by including the following term in equation \eref{atom_eqn}:
\begin{align}
&-i \frac{\hbar}{2} K_{3}  |\phiax |^{4}\phiax.
\end{align} 
As the next step we derive the evolution equation for $G_{N/A}$ from the Heisenberg equations for the operators $\hat{\chi}^{\dagger}(x')\hat{\chi}(x)$ and $\hat{\chi}(x')\hat{\chi}(x)$ which gives us:
\begin{align}
\nonumber
&i \hbar  \frac{\partial \gax}{\partial t}=
\langle \big[\hat{\chi}(x')\hat{\chi}(x) , \hat{H}\big]\rangle=
\nnl
&\big( \hamax + \hamaxs \big) \gax +2 U_{o} \big( \mods{\phiax}  + \mods{\phiaxs} 
\nnl
&+ \gnxx +G_{N}\left( \mathbf{x'},\mathbf{x'} \right) \big) \gax 
\nnl
&+ U_{o} \big(  \phiax^{2} \gnxstar  +\phiaxs^{2} \gnx 
\nnl
&+ \gaxx  \gnxstar   + G_{A}\left( \mathbf{x'},\mathbf{x'}\right) \gnx \big) 
\nl
&+ U_{o} \big(\phiax^{2} + \gaxx \big)\deltf,
\label{ga_eqn}
\\[0.5cm]
\nonumber
&i \hbar  \frac{\partial \gnx}{\partial t}=
\langle \big[\hat{\chi}^{\dagger}(x')\hat{\chi}(x), \hat{H}\big]\rangle=
\nnl
&\big( \hamax - \hamaxs \big) \gnx +2 U_{o} \big( \mods{\phiax}  - \mods{\phiaxs}
\nnl
&+ \gnxx - G_{N}\left( \mathbf{x'},\mathbf{x'} \right)  \big) \gnx 
\nnl
 &+ U_{o} \big( \gaxx \gaxstar- G_{A}^{*}\left( \mathbf{x'},\mathbf{x'} \right)\gax \big)
\nl
&+U_{o} \big(\phiax^{2}\gaxstar -\phicaxs^{2} \gax \big).
\label{gn_eqn}
\end{align}
Equations (\ref{atom_eqn})-(\ref{gn_eqn}) constitute the so called dynamical HFB equations and are identical to those of \rref{holland:burst}, without the molecular field (see below). The calculation of quantum corrections in a local field theory naturally leads to divergences. Therefore a renormalization of the theory is necessary. As the renormalization condition we demand that the amplitude for low energy atom scattering is given by $U=4\pi \hbar^{2} a/m$ where $a$ is the s-wave scattering length \cite{book:pethik}. Ultraviolet (high momentum) divergences that arise when this amplitude is determined from the Hamiltonian (\ref{Hamiltonian}) are regularized by a momentum cut-off $K$. This yields a relation between the physical interaction strength $U$ and the parameter $U_{o}$ in the Hamiltonian. Following Kokkelmans \textit{et al.} \cite{holland:renorm}: 

\begin{align}
U_{o}&=\dfrac{U}{1-\alpha U},&&
\alpha=\dfrac{m K}{2 \pi^{2} \hbar^{2}}.
\label{renormalization}
\end{align} 
%
%:
This process also provides a natural interpretation of the delta function in \eref{ga_eqn} in terms of its truncated Fourier transform. Some subtle issues connected to the renormalization procedure will be discussed in the final section of this paper. The reason that quantum field effects might improve the agreement in \fref{fig:collapsetime} is the statistical factor of two in the interaction between the atomic mean field and the uncondensed component in \eref{atom_eqn}. If a sufficient number of atoms are transferred to uncondensed modes during collapse, it is expected to occur earlier. 

%This interaction is not properly renormalized \cite{duinestoof:long} but should be a sufficient approximation to investigate whether excitations have an effect on the collapse time;

\subsection{Resonance theory}
\label{section_molecules}

In the vicinity of a Feshbach resonance the atomic interactions are strongly energy dependent.  To investigate effects of this energy dependence it is possible to model the Feshbach resonant interaction by including a molecular field $\psimx$ in the model \cite{holland:burst}. The additional terms in the Hamiltonian are then:

\begin{align}
& \intas \psiadmx
\left(-\frac{\hbar^2}{4m} \nabla_\mathbf{x}^2 +2V(\mathbf{x}) -\nu_{o} \right) \psimx
\CR
& + \frac{g_{o}}{2} \intas \psiadmx \psiax \psiax +H.c.
\label{mol_hamil}
\end{align}
and the renormalization formalism presented in \cite{holland:renorm} now connects the atom-molecule coupling $g$, molecular detuning $\nu$ and background atom-atom scattering $\sub{U}{bg}$ to the effective interaction $\sub{U}{eff}$ that the atoms experience due to their coupling to the molecular field by:
\begin{align}
U_{o}&=\dfrac{U_{bg}}{1-\alpha \: U_{bg}},
\\
g_{o}&=\dfrac{g}{1- \alpha \:U_{bg}},
\\
\nu_{o}&=\nu- \alpha \dfrac{g g_{o}}{2}=\nu- \alpha \dfrac{g^{2}}{2(1-\alpha \:U_{bg})},
\\
\sub{U}{eff}&=\sub{U}{bg}\left( 1 +  \frac{g^2}{2 \sub{U}{bg} \nu} \right).  
\end{align}
Note that the parameters $\sub{U}{bg},g,\nu$ are physical parameters, distinct from the bare parameters $U_{o},g_{o},\nu_{o}$, which are just the parameters in the Hamiltonian and depend on the chosen cut-off. The atom-molecule coupling is related to the width of the Feshbach resonance $\kappa$ by:
\begin{align}
g&=\sqrt{\kappa \sub{U}{bg}}=\sqrt{\Gamma \Delta  \sub{\mu}{bg}}\hspace{0.2cm}  ,
\end{align}
where $\Gamma$ is the width in terms of the magnetic field \cite{jila:revision}, and $ \Delta  \sub{\mu}{bg}$ is the difference in magnetic moments of the open and closed channels \cite{holland:deltamu}. The coupling is therefore experimentally determined, as is the background scattering length $\sub{U}{bg}$ \cite{jila:revision}. 

In deriving the equations of motion for the coupled atom-molecule system quantum fluctuations of the molecular field are neglected, that is we assume $\langle \psimx \rangle = \phimx$ which yields:
%since already the molecular population is expected to be small.
%The equations of motion for the coupled atom-molecule system, assuming $\langle \psimx \rangle = \phimx$ are now:
%
\begin{align}
&i \hbar  \frac{\partial \phiax}{\partial t}=\left(-\frac{\hbar^2}{2m} \nabla_\mathbf{x}^2 +V(\mathbf{x}) 
 +U_{o} |\phiax |^{2} \right) \phiax
\CR
&+ 2 U_{o} \gnxx  \phiax +U_{o} \gaxx \phicax 
\CR
&+g_{o} \phimx \phicax,
\label{atom_eqn_mol}
\\
\CR
&i \hbar  \frac{\partial \phimx}{\partial t}=\left(-\frac{\hbar^2}{4m} \nabla_\mathbf{x}^2 + 2V(\mathbf{x}) -\nu_{o} \right)  \phimx 
\CR
&+\frac{g_{o}}{2}\left(\gaxx +\phiax^{2} \right).
\label{mol:eqn}
\end{align}
%
%Quantum fluctuations of the molecular field were neglected since already the molecular population is expected to be small. 
For the modifications of the remaining two Hartree-Fock Bogoliubov equations (\ref{ga_eqn}-\ref{gn_eqn}) by the coupling to molecules we refer to \rref{holland:burst}.

\section{Results}
\subsection{HFB Approach}
\label{section_hfbresults}

While the GPE can be tackled even in a completely asymmetric situation \cite{savage:coll}, doing so for the set of equations (\ref{atom_eqn})-(\ref{gn_eqn}) even with cylindrical symmetry would present a serious numerical problem since the dimensionality of the correlation functions would only reduce from six to five. The resulting requirements for memory and computation time would exceed those for spherical geometry by two orders of magnitude. 
One problem is, that the violent nature of the collapse and the resulting spreading of the  momentum space wavefunction do not allow radial grids smaller than 128 points, 
%If all the angular structure appearing in the full cylindrical description can be expanded in some nontrivial basis set, 
and our experience is that the minimum number of grid points to describe an angular variable lies around 8. Higher numbers are preferable. 
We are therefore forced to model the experiment with a spherically symmetric model. The trap frequency is chosen to equal the geometric mean of the experimental trap frequencies: $\omega = (\omega_{\bot}^{2} \omega_{\|} )^{1/3}$. With $\omega_{\bot}=$17.5 Hz and  $\omega_{\|}=$6.8 Hz from \cite{jila:nova}, $\omega=$12.8~Hz. To determine the influence of geometry on the collapse, we numerically solved the spherically symmetric GPE for this trap parameter. The results are included in \fref{fig:collapsetime} and show that the spherical situation gives quite a good estimate of the complete three dimensional GPE collapse time.
In order to simplify equations (\ref{ga_eqn}) and (\ref{gn_eqn}) for the case of spherical symmetry we work in coordinates:
\begin{align}
 &R=|\mathbf{x}|, \hspace{0.5cm} R'=|\mathbf{x'}|,  \hspace{0.5cm} \beta=\cos{\theta},
 \end{align}
where $\theta$ is the angle between $\mathbf{x}$ and $\mathbf{x'}$. Due to the restriction to spherical symmetry the correlation functions will not depend on the overall orientation of $\mathbf{x}$ and $\mathbf{x'}$, or on the azimuthal angle of $\mathbf{x'}$ around $\mathbf{x}$, reducing the dimensionality of $G_{A}$ and $G_{N}$ from six to three. Following \cite{holland:burst} the $\beta$ dependence is expanded in terms of Legendre polynomials $P_{n}(\beta)$. Numerically, we model the delta function in \eref{ga_eqn} by a step function on one spatial grid point. The corresponding cut-off in \eref{renormalization} is thus $K=\pi/ \Delta R$, where $\Delta R$ is our grid spacing. Since we calculate spatial derivatives with the FFT all radial grids need to incorporate an unphysical negative range.
We now define the functions:
\begin{align}
&\phi_{a}(R)=\frac{\phiar}{R}, 
&&
\tilde{G}_{N/A}(R,R',\beta)=\frac{G_{N/A}(R,R',\beta)}{R R'},
\end{align}
and expand the correlation functions as:
\begin{align}
\gnpol&= \sum_{n=0}^{M-1} \gnpolm{n} P_{n}(\beta),
\\
\gapol&= \sum_{n=0}^{M-1} \gapolm{n} P_{n}(\beta),
 \end{align}
where $M$ is the number of Legendre Polynomials employed. We use the additional notation:
\begin{align}
\bar{G}_{N/A}(R)&\equiv\tilde{G}_{N/A}(\mathbf{R},\mathbf{R})= \sum_{n=0}^{M-1}\tilde{G}_{N/A}^{(n)}(R,R).
\end{align}
The final form of our equations reads:
\begin{align}
\nonumber
&i \hbar  \frac{\partial \phiar}{\partial t}=
\nnl
&\left(-\frac{\hbar^{2}}{2m} \frac{\partial^{2}}{\partial R^{2}}  +V(R)
- i \frac{\hbar}{2} \frac{K_{3}}{R^4}  |\phiar |^{4} \right) \phiar 
\nnl
&+ \frac{U_o}{R^2} \left(2 \gndiag + |\phiar |^{2} \right) \phiar 
\\
&+ \frac{U_o}{R^2} \gadiag \phiarstar  , 
\label{atom_fin}
\end{align}
for the atomic condensate. The expansion coefficients obey the following equations of motion:
\begin{align}
\nonumber
&i \hbar  \frac{\partial \gnpolm{n}  }{\partial t}=
\nnl
&\bigg[-\frac{\hbar^{2}}{2m} \left(
       \frac{\partial^{2}}{\partial R^{2}} 
    -  \frac{\partial^{2}}{\partial R'^{2}}
      -n(n+1)\left(\frac{1}{R^{2}} -\frac{1}{R'^{2}} \right) \right) 
\nnl
&+ V(R) -V(R') +2 U_o  \frac{ |\phiar |^2 + \gndiag}{R^2}
\nnl
 &-2 U_o \frac{ |\phiard |^2 +\gndiags}{R'^2}  \bigg] \gnpolm{n}
\nnl
&+ U_{o} \frac{\phiar^{2} + \gadiag}{R^{2}}\gapolmstar{n}
\\
\label{gn_fin}
&-U_{o}\frac{\tilde{\phi}_{a}^{2 *}(R')+\gadiagsstar}{R'^{2}}\gapolm{n},
\end{align}
\begin{align}
\nonumber
&i \hbar  \frac{\partial \gapolm{n}}{\partial t}=  
\nnl
 &\bigg[ -\frac{\hbar^{2}}{2m} \left(
       \frac{\partial^{2}}{\partial R^{2}} 
    +  \frac{\partial^{2}}{\partial R'^{2}} 
      -n(n+1)\left(\frac{1}{R^{2}} +\frac{1}{R'^{2}} \right)\right) 
\nnl
&+V(R) +V(R')  +2 U_o  \frac{ |\phiar |^2 +  \gndiag}{R^2} 
 \nnl
 & +2 U_o  \frac{ |\phiard |^2 +  \gndiags}{R'^2} \bigg] \gapolm{n}
\nnl
&+ U_{o} \frac{\phiar^{2} + \gadiag}{R^{2}}\gnpolmstar{n}
\nnl
 &+U_{o}  \frac{\phiard^{2} + \gadiags}{R'^{2}}\gnpolm{n}
 \\
\label{ga_fin}
&+  U_{o}\left(\phiar^{2}+\gadiag \right) \frac{2n+1}{4 \pi R^{2}}\delta(R-R').
\end{align}

As initial checks of our code we reproduced the results of Milstein \textit{et al.} \cite{holland:burst} and Holland \textit{et al.} \cite{holland:homog}, using the molecular field and the equations of \rref{holland:burst}. We then computed a solution to Eqs.~(\ref{atom_fin})-(\ref{ga_fin}), without the molecular field for the experimental parameters \cite{jila:nova} of a case (ii) collapse ($\sub{a}{collapse}=-12a_{o}$), starting with a gaussian initial state and no uncondensed atoms due to the absence of interactions ($\sub{a}{init}=0$).

\begin{figure}
\epsfig{file={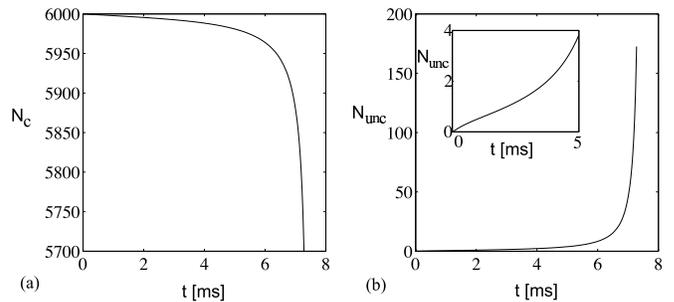},width=\figuresize}
\caption{Number of condensed (a) and uncondensed (b) atoms during a case (ii) collapse with $\sub{a}{collapse}=-12 a_{o}$ until $t=7.3$ ms. The initial atom number was $N_{o}=6000$, all in the condensate. A three-body loss term with $K_3 = 1 \times 10^{-27}$ cm$^6$s$^{-1}$ is included for the condensate. The GPE value for $\sub{t}{collapse}(a/a_{o}=-12)$ is $7.45$ ms. Numerical requirements prevent the simulations from reaching the exact GPE collapse point. Inset of (b): initial evolution of uncondensed-atom number.}
\label{fig:uncondensed}
\end{figure}

We found that the BEC does not collapse earlier than in the corresponding spherical GPE simulations. The number of atoms in excitation modes, shown in \fref{fig:uncondensed}, does not grow fast enough to accelerate the collapse. Only just before the collapse time predicted by the GPE are large numbers of uncondensed atoms created.
%, and therefore are a candidate for the burst atoms observed in the experiment. 
We compared our generated numbers of uncondensed atoms with the predictions of Yurovsky \cite{yurovsky:instabilities} %taking into account our initially Gaussian density distribution but otherwise neglecting any effects of the inhomogeneous situation. 
for the central, approximately homogeneous region of a large condensate ($N_{init}=50000$, $a_{init}=50a_{o}$).
For the case of $\sub{a}{collapse}=-12a_{o}$ and times much smaller than the nonlinear interaction time \cite{yurovsky:instabilities} $\sub{t}{NL}=(U |\phi(0)|^{2} /\hbar)^{-1}\approx 4$ ms we found that the theories agree.
%to within 50\%, even under this crude approximation.

\begin{figure}
\centering
\epsfig{file={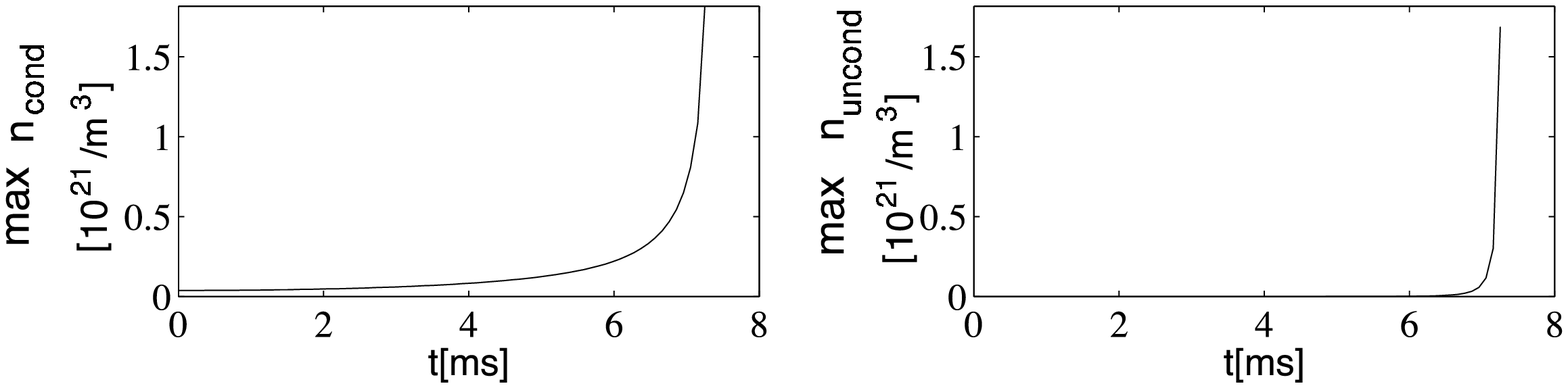},width=\figuresize}
\\
\epsfig{file={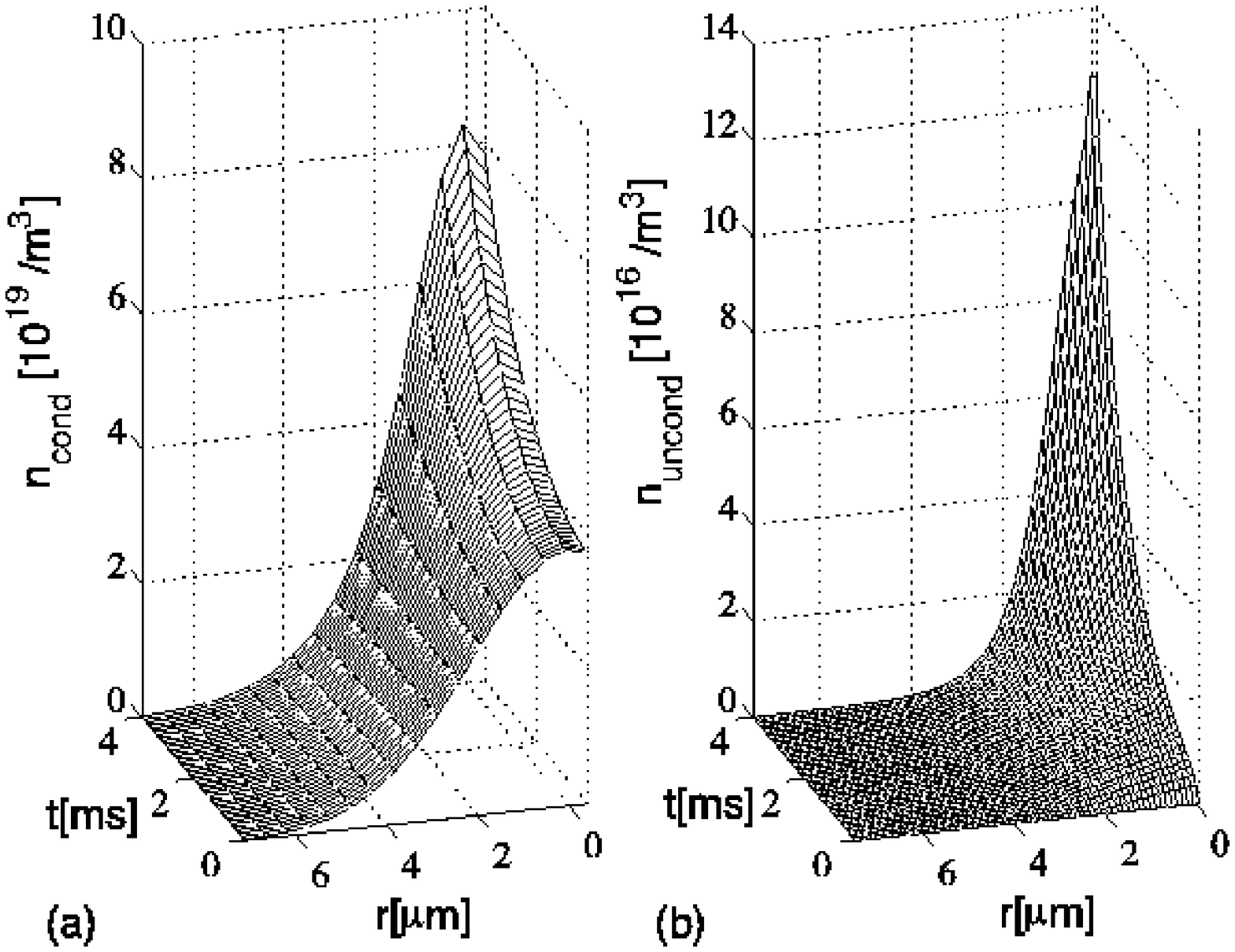},width=\figuresize}
%\epsfig{file={densities.eps},width=10cm}
%:
\caption{a) Time evolution of condensate density $\sub{n}{cond}=|\phi|^{2}$ and b) density of uncondensed atoms $\sub{n}{uncond}=\bar{G}_{N}$ for a case (ii) collapse with $\sub{a}{collapse}=-12 a_{o}$. Above we display the corresponding peak densities until $t=7.3$ ms.}
\label{fig:densities}
\end{figure}

\begin{figure}
%:
\epsfig{file={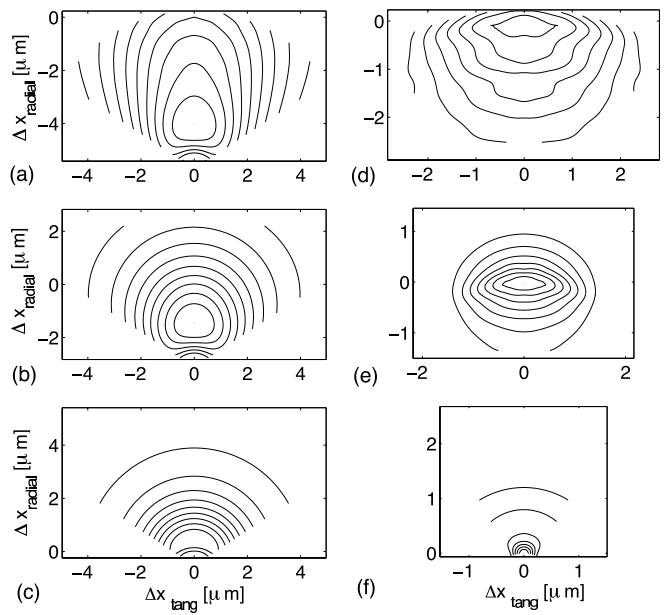},width=\figuresize}
%\epsfig{file={correl_contour.ps},width=15cm}
%\put(-265,360){\large a)}
\caption{Equal amplitude contours of correlation functions $G_{N/A}(\mathbf{R_{0}},\mathbf{R})$ for fixed $\mathbf{R_{0}}$. We plot correlations for $\mathbf{R}$ varying in a plane that contains the trap centre and the point $\mathbf{R}_{0}$. The origin of cartesian coordinates ($\Delta \sub{x}{tang}$, $\Delta \sub{x}{radial}$) in this plane is taken to be $\mathbf{R_{0}}$, a distance $|\mathbf{R_{0}}|=R_{o}$ from the trap centre. The $\Delta \sub{x}{radial}$ axis connects $\mathbf{R_{0}}$ with the trap centre, which is thus located at the bottom-centre of each plot. The contours are equally spaced between the maximum value and 1/10 (3/10) of it, for $G_{N}$ ($G_{A}$). 
%Data points stem from a $128 \times 128 \times 128$ grid in (R,R',$\theta$) but were interpolated to four times higher resolution. 
Normal density $G_{N}$ around a) $R_{0}=5.8$ $\mu$m, b) $R_{0}=3.2$ $\mu$m, c) $R_{0}=0.6$ $\mu$m. Anomalous density $G_{A}$ around d) $R_{0}=5.8$ $\mu$m, e) $R_{0}=3.2$ $\mu$m, f) $R_{0}=0.6$ $\mu$m. The snapshot is taken during a case (ii) collapse with $a=-12a_{0}$ at $\sub{\tau}{evolve}=5$ ms.}
\label{fig:correl}
\end{figure}

Since the main focus of this paper is the investigation of $\sub{t}{collapse}$, we did not compute an HFB solution beyond the collapse point, where the numerical cost would have become immense. The main effort went into the computation of the correlation functions of the condensate fluctuations $G_{N}$ and $G_{A}$ from which the uncondensed density $\bar{G}_{N}$ is derived. 
\fref{fig:densities} shows the initial evolution of the densities of the condensed and uncondensed atoms and the corresponding peak densities for the simulated period of time. 
To illustrate the structure of the complete correlation functions $G_{N/A}(\mathbf{R},\mathbf{R'})$ some representative samples are displayed in \fref{fig:correl}.

The results presented were computed on a grid with $N_r=256$ points for $R$ and $R'$, ranging from -15 $\mu$m to +15 $\mu$m each, and eight Legendre polynomials. We checked that the results do not vary for changed spatial grid size or number of Legendre polynomials. The timesteps were 50 ns and 100 ns to check convergence. The variation of $N_{r}$ also verified the cut-off independence of our simulations. Our multiprocessor code was run on 8 or 16 CPUs \cite{APAC} and employed the RK4IP algorithm developed by the BEC theory group of R. Ballagh at the University of Otago \cite{RK4IP}. Parts of the simulations were done with the aid of the high level language XMDS \cite{XMDS}.

\subsection{Resonance theory}

We also checked that the implementation of the molecular field as described in \sref{section_molecules} (thus an energy-dependent model of scattering properties) does not change the main result of this paper; that condensate depletion is not sufficient to significantly accelerate the collapse. If we want to apply the resonance theory formalism \cite{holland:burst} to the exact experimental situation, the initial state for the molecular field needs to be carefully adjusted. 
Solving the molecular \eref{mol:eqn} approximately, far away from resonance (a condition fulfilled for the collapse-scenarios concerned), yields \cite{calz:hu}:
\begin{align}
 \phi_{m}(\mathbf{x},t_{o})&=\frac{g}{2 \nu}\phi^{2}_{a}(\mathbf{x},t_{o}).
\end{align}
Simulations done with this particular molecular initial state show that the molecular field performs rapid oscillations around the instantaneous value of $g/(2\nu) \phi^{2}_{a}(\mathbf{x},t)$. 
The effect on the interaction terms in the modified GP equation (\ref{atom_eqn_mol}) is just the generation of an effective atom-atom interaction of strength
%as if we had replaced $\sub{U}{bg}$ by $\sub{U}{eff}$ in a simulation that contains only the atom field
$\sub{U}{eff}$, supporting the view presented in \cite{calz:hu}, that far off resonance the molecular field is not important for the condensate evolution. Due to the large atom-molecule coupling, $g \phi_{m}$ is of the order of $U |\phi_{a}|^{2}$ even for the very small molecular population that we observed. In order to make predictions with the same precision as those presented in \sref{section_hfbresults}, it might therefore be necessary to include quantum fluctuations of the molecular field as well. Preliminary results neglecting these do not indicate production rates of uncondensed atoms that differ qualitatively from those in \fref{fig:uncondensed}. Nonetheless a more careful implementation of the molecular field might be a subject for further studies.


\section{Discussion}

After our investigations an open question remains: what is the correct quantitative theory that describes the case (ii) collapse experiment? We therefore discuss some limitations of our approach, and possible improvements and alternatives. 

First it might be possible that a quantum treatment with the proper experimental cylindrical symmetry yields agreement, but it is not obvious how a changed symmetry would cause a qualitative change in the low production rates of uncondensed atoms that we found.

Another factor that might be relevant is any uncondensed population in the initial state, which we assumed to be zero. But preliminary simulations that include an initial thermal cloud show only a small acceleration of the collapse (of the order of 5\%) for the experimental temperature of 3 nK. Here we estimate the collapse time from the onset of rapid decline of the condensate population. For temperatures as high as 7 nK the collapse time is about 10\% shorter \cite{davis:wigner}. But even this cannot by itself account for the discrepancy between theoretical and experimental data. 
%If the ramp from the evaporation scattering length $a=240a_{o}$ to $a=0$ in the experiment was not perfectly adiabatic with respect to the interatomic interaction, some fraction of the condensate depletion could have remained. 

A drawback of the HFB method is that the interaction between uncondensed and condensed atoms is not properly renormalized \cite{duinestoof:long}, since this would require the inclusion of higher order correlation functions \cite{duinestoof:long,burnett:ghfb}. Since the effects of renormalization on the coupling constant are small in the parameter regime of case (ii), it is not a priori clear whether a more complete renormalization would significantly alter the number of uncondensed atoms. We applied the gapless HFB method of \cite{burnett:ghfb} to a case (ii) collapse and found no changes in our results.

A further step could be to employ a different factorisation scheme for the correlation functions, for example the method of non-commutative cumulants put forward in \cite{burnett:cumulants} which differs from ours in its truncation method. To determine the rate of spontaneous production of uncondensed atoms in that formalism we would need to include the higher order correlation functions in our simulations.

Finally, the growth of the uncondensed fraction by elastic collisions as investigated in \cite{duinestoof:evaporation,duinestoof:feshbachdynamics} might also play an important role in a case (ii) collapse. Its physics seems not to be described by the HFB equations since the creation of pairs can be identified as the leading contribution to the production of uncondensed atoms in the limit of weak coupling $U_{o}$. This does not mean that relevant physics is missing in the HFB method, as a coherent elastic collision process with one atom scattering into a previously uncondensed mode might transfer coherence into that region of Hilbert space \cite{burnett:projGPE}. In that case the above process would not populate $G_{N}$ since we define the quantum field $\hat{\chi}$ to have zero expectation value. 
 
%Whether these elastic collisions require a separate treatment might therefore hinge on the actual definition of the condensate employed.
%:

%\section{acknowledgments}
\acknowledgments
We would like to thank M. Holland for providing the code used in \cite{holland:burst} for the purpose of comparison. We also thank R. Duine, M. Davis and B. Blakie for discussions, as well as S. M\'etens for his comments. This research was supported by the Australian Research Council under the Centre of Excellence for Quantum-Atom Optics and by an award under the Merit Allocation Scheme of the National Facility of the Australian Partnership for Advanced Computing. 
%\end{acknowledgments}

%\bibliography{bosenova2}

\end{document}